\documentclass[letter,twocolumn]{jpsj3}
\usepackage{txfonts}

\title{$^{153}$Eu and $^{69, 71}$Ga Zero-Field  NMR Study of Antiferromagnetic State in EuGa$_{4}$}

\author{
\name{Mamoru \surname{Yogi}}$^1$\thanks{myogi@sci.u-ryukyu.ac.jp},
\name{Saori \surname{Nakamura}}$^1$,
\name{Nonoka \surname{Higa}}$^1$,
\name{Haruo \surname{Niki}}$^1$,
\name{Yusuke \surname{Hirose}}$^2$,
\name{Yoshichika \surname{\={O}nuki}}$^2$, and
\name{Hisatomo \surname{Harima}}$^3$
}

\inst{
$^1$Department of Physics and Earth Sciences, Faculty of Science, University of the Ryukyus, Okinawa 903-0213, Japan\\
$^2$Graduate School of Science, Osaka University, Toyonaka 560-0043, Japan \\
$^3$Graduate School of Science, Kobe University, Nada-ku, Kobe 657-8501, Japan
} 

\abst{
We report $^{153}$Eu and $^{69,71}$Ga NMR under a zero magnetic field on the antiferromagnetic state of EuGa$_{4}$ with $T_{\rm{N}}\approx 16$ K. 
We have successfully observed a $^{153}$Eu zero-field NMR signal with well-resolved nuclear quadrupole splitting in the antiferromagnetic state of EuGa$_{4}$.
$^{69,71}$Ga zero-field NMR spectra were also observed below $T_{\rm{N}}$.
The internal field and nuclear quadrupole frequency are estimated from a simulation of the spectra by the exact diagonalization of the nuclear spin Hamiltonian matrix. 
The asymmetrically split zero-field NMR spectra were explained by considering a configuration of the magnetic moments of Eu$^{2+}$ lying in the basal $ab$-plane.
The temperature dependence of the internal field, which is proportional to the sublattice magnetization, can be explained by the Brillouin function with $J=S=7/2$.
}

\kword{europium compound, EuGa$_{4}$, antiferromagnetism, zero-field NMR}

\begin{document}
\maketitle

Strong electron correlations in rare-earth-based compounds induce various interesting physical phenomena such as heavy-electron behavior,  multipole order, unconventional superconductivity, spin or valence quantum critical fluctuations, and non-Fermi liquid behavior.\cite{Onuki_HF,Kuramoto_MO,Watanabe_VQ}
Eu is a rare-earth element known to have two kinds of valence states: Eu$^{2+}$ ($4f^7$) and Eu$^{3+}$ ($4f^6$).
The divalent Eu state is magnetic ($J=S=7/2$, $L=0$), where $J$ is the total angular momentum, $S$ is the spin angular momentum, and $L$ is the orbital angular momentum.
Therefore, the compounds with divalent Eu ions tend to order magnetically, following the Ruderman-Kittel-Kasuya-Yosida (RKKY) interaction.
In contrast, the trivalent Eu state is non-magnetic ($J=0$, $S=L=3$).
Note that the valence of Eu in some compounds is changed by temperature, magnetic field, and pressure.\cite{Hesse_EuNi2Ge2,Wortmann_EuNi2Si2,Matsubayashi_EuFe2As2,Mitsuda_EuRh2Si2}
A mixed valence state is also an interesting feature of the compound.

\begin{figure}[tb]
  \begin{center}
    \includegraphics[keepaspectratio=true,width=60mm]{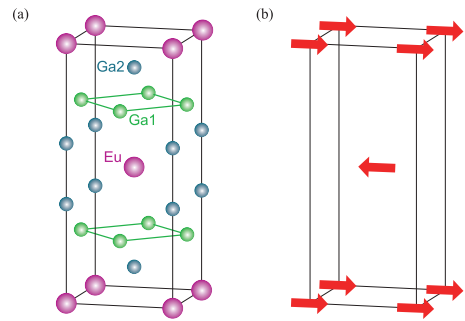}
  \end{center}
  \caption{(Color online) (a) Crystal structure of EuGa$_{4}$. (b) Magnetic moments configuration assuming the antiferromagnetic structure type-I (see text).}\label{structure}
\end{figure}

In this study, we focus on the Eu intermetallic compound EuGa$_{4}$, which crystallizes in the BaAl$_{4}$-type tetragonal structure (space group: No.139, $I4/mmm$) shown in Fig. \ref{structure}(a).\cite{Bobev_EuGa4,Nakamura_EuGa4}
Eu atoms occupy the corners and the center of the body-centered lattice with local symmetry ($4/mmm$ in Hermann-Mauguin notation or International notation).
As for Ga atoms, they have two crystallographically inequivalent sites, denoted Ga1 ($\bar{4}m2$) and Ga2 ($4mm$). 
A divalent Eu state was demonstrated from an isotropic behavior of the magnetic susceptibility in the paramagnetic state with the effective magnetic moment $\mu_{\rm eff} = 7.86$ $\mu_{\rm B}/$Eu.\cite{Nakamura_EuGa4}
This is close to a divalent value of $7.94$ $\mu_{\rm B}/$Eu.
Antiferromagnetic (AFM) order was confirmed from the measurements of resistivity, magnetic susceptibility, specific heat, and thermoelectric power below a N\'{e}el temperature $T_{\rm N} \approx 16$ K.\cite{Bobev_EuGa4,Nakamura_EuGa4}
In the AFM state, the magnetic susceptibility for $H\parallel [100]$ and [110] decreases with decreasing temperature, while the susceptibility for $H\parallel [001]$ is almost unchanged; thus, the magnetic moments are considered to lie in the $ab$-plane.\cite{Nakamura_EuGa4}
Moreover, a recent neutron scattering experiment has clarified that the AFM structure is of type-I. \cite{Kawasaki_EuGa4}
The possible magnetic moment configuration in the AFM state is described by the arrows in Fig. \ref{structure}(b).
In addition, the possible emergence of the charge density wave (CDW) order was also reported from the measurements of electrical resistivity and thermoelectric power under pressure.\cite{Nakamura_EuGa4}

NMR measurement is a powerful technique for investigating magnetic properties from the microscopic viewpoint.
The nuclear properties of both Eu and Ga nuclei are listed in Table \ref{Nuclear}.\cite{NMRprop}
The NMR of both nuclei is basically possible in EuGa$_{4}$.
However, a ligand nucleus is usually used in pulse NMR experiments on rare-earth-based magnetic compounds because the relaxation time of rare-earth nuclei is too short for observing the NMR signal.
As for divalent Eu compounds, to the best of our knowledge, the signal of the Eu nucleus is only observed in a magnetically ordered state of Eu chalcogenide compounds Eu$X$ ($X$ = O, S, Se, and Te).\cite{Fekete_EuO,Heller_EuS,Komaru_EuSe,Hihara_EuTe}
In this circumstance, we have attempted NMR measurement of the Eu nucleus in EuGa$_{4}$, and successfully observed a $^{153}$Eu NMR signal in the AFM state.
In this letter, we report on the magnetic property of EuGa$_{4}$ in the AFM state determined through $^{153}$Eu and $^{69, 71}$Ga NMR measurements under a zero magnetic field.

\begin{table}
\caption{Data of Eu and Ga isotopes: nuclear spins $I$, nuclear gyromagnetic ratios $\gamma_n$, nuclear quadrupole moments $Q$, and natural abundances N.A.\cite{NMRprop}}
\label{Nuclear}
\begin{center}
\begin{tabular}{ccccc}
\hline
 & $I$ & $\gamma_n/2\pi$ (MHz/T) & $Q$ ($10^{-26}$cm$^2$) & N.A. (\%) \\
\hline
$^{151}$Eu & 5/2 & 10.5854 & 90.3 & 47.81 \\
$^{153}$Eu & 5/2 & 4.6744 & 241.2 & 52.19 \\
$^{69}$Ga & 3/2 & 10.2475 & 17.1 & 60.108 \\
$^{71}$Ga & 3/2 & 13.0204 & 10.7 & 39.892 \\
\hline
\end{tabular}
\end{center}
\end{table}

Single crystals of EuGa$_{4}$ were grown by the Ga self-flux method.
Details of the sample preparation are described elsewhere.\cite{Bobev_EuGa4,Nakamura_EuGa4}
The crystals were powdered to facilitate applied rf-field penetration.
The $^{153}$Eu and $^{69,71}$Ga NMR measurements were performed by a conventional spin-echo method using a conventional phase-coherent pulsed spectrometer in the range of $T = 1.6-7.5$ K for $^{153}$Eu and $T = 1.5-14$ K for $^{69,71}$Ga.
The NMR spectra were obtained by sweeping the frequency and integrating the spin-echo signal intensity step by step.

\begin{figure}[tb]
  \begin{center}
    \includegraphics[keepaspectratio=true,width=75mm]{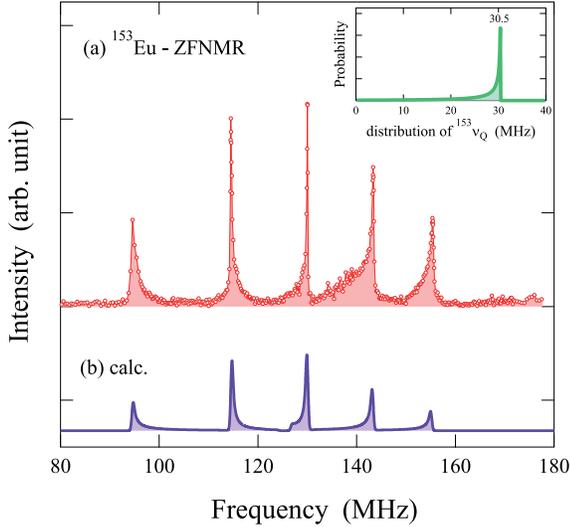}
  \end{center}
  \caption{(Color online) (a) $^{153}$Eu-ZFNMR spectrum at $T=4.2$ K. (b) Calculated spectrum for $^{153}$Eu in the case of $H_{\rm{int}}^{\rm{Eu}}\perp V_{zz}^{\rm{Eu}}$ assuming a log-normal distribution of $^{153}\nu_{Q}$ as shown in the inset.}\label{Eu}
\end{figure}
Generally, an external magnetic field is necessary for the observation of NMR.
However, if there is a large internal field transferred on the nuclei from adjacent magnetic moments in the magnetically ordered state, the Zeeman interaction between the internal field and the nuclear magnetic moment causes the level splitting of the nuclear spin states.
Then, we can observe the NMR signal under zero magnetic field, which is called the zero-field NMR (ZFNMR).

Figure \ref{Eu}(a) shows the $^{153}$Eu-ZFNMR spectrum at $T=4.2$ K.
$^{153}$Eu has a nuclear spin $I=5/2$; therefore, five split peaks were observed when there is  a finite electric field gradient (EFG).
Note that the frequency separations between neighboring peaks are different, which gives us important information on the direction of the internal field.
The Eu site in EuGa$_{4}$ has a tetragonal local symmetry ($4/mmm$); hence, the EFG becomes axially symmetric, namely, the asymmetry parameter of the EFG becomes zero ($\eta = 0$).
In addition, the main principal axis of the EFG, denoted $V_{zz}$, is parallel to the $c$-axis.
Thus, the nuclear spin Hamiltonian in the AFM state is given by 
\begin{equation}
\mathcal{H} = -\gamma_{n} \hbar \mib{I} \cdot \mib{H}_{\rm{int}} + \frac{h \nu_{Q}}{6}\left[  3I_{z}^2 - I^2 \right].\label{Hami}
\end{equation}
The first term of the Hamiltonian represents the Zeeman interaction between the nuclear magnetic moment $\mib{\mu}_{n} = \gamma_{n} \hbar \mib{I}$ and the internal field $\mib{H}_{\rm{int}}$, where $\gamma_{n}$ is the nuclear gyromagnetic ratio and $\mib{I}$ is the nuclear spin.
The second term in the Hamiltonian represents the nuclear quadrupole interaction between the EFG and the nuclear quadrupole moment $Q$.
Here, $\nu_{Q}$ is the nuclear quadrupole frequency defined by $\nu_{Q}\equiv 3eQV_{zz}/2I(2I-1)h$.
If $H_{\rm{int}}\parallel V_{zz}$ and the Zeeman interaction is larger than the nuclear quadrupole interaction, the separations between peaks become the same.
On the other hand, if $H_{\rm{int}}\perp V_{zz}$, the separations of the peak become inequivalent, which is consistent with the observed $^{153}$Eu-ZFNMR spectrum, indicating ordered magnetic moments lying in the $ab$-plane.
This is in good agreement with the results of magnetic susceptibility and neutron diffraction experiments.

The numerical calculation of the NMR resonance frequencies by the exact diagonalization of the nuclear spin Hamiltonian matrix explains well the peak positions of the spectrum.
From this calculation, an internal field at the Eu nucleus $H_{\rm{int}}^{\rm{Eu}} = 27.08$ T and $^{153}\nu_{Q} = 30.5$ MHz are obtained at 4.2 K; the  $H_{\rm{int}}^{\rm{Eu}}$ obtained is close to the value of 26.6 T determined by M\"{o}ssbauer spectroscopy.\cite{Vries_EuGa4}
A broad tail of each peak indicates a microscopic inhomogeneity of the electric state at the Eu site.
This asymmetric shape of the spectrum can be explained by considering a small distribution of the EFG.
Figure \ref{Eu}(b) shows a calculated spectrum assuming a log-normal distribution of $^{153}\nu_{Q}$, as shown in the inset of Fig. \ref{Eu}.
The calculation well explains the observed $^{153}$Eu-ZFNMR spectrum.
We infer two possible origins of this distribution.
One is an intrinsic origin.
An anomaly related to the CDW order was reported from the thermoelectric power measurement at ambient pressure.\cite{Nakamura_EuGa4}
Therefore, a short range order would occur at ambient pressure, causing a small distribution of the EFG.
The other is an extrinsic origin.
The powdering of the crystal might cause microscopic distortions to the sample, causing a distribution of the EFG.

Eu has two isotopes, $^{151}$Eu and $^{153}$Eu, as listed in Table. \ref{Nuclear}.
Using $H_{\rm{int}}^{\rm{Eu}} = 27.08$ T, we infer that the $^{151}$Eu-ZFNMR spectrum appears at approximately $270\sim 300$ MHz.
However, we could not observe the signal.
This is because the relaxation time of $^{151}$Eu is shorter than that of $^{153}$Eu.
If a magnetic fluctuation is a predominant relaxation process, the relaxation rate is proportional to $\gamma_n^2$.
Therefore, the relaxation time of $^{151}$Eu becomes 5 ($\approx (^{151}\gamma_n /^{153}\gamma_n )^2$) times shorter than that of $^{153}$Eu.
We used the shortest possible $\tau$ of 10 $\mu$s for $^{153}$Eu-ZFNMR, where $\tau$ is the time between the excitation pulse and the refocusing pulse, implying that $\tau = 2$ $\mu$s is necessary for observing the $^{151}$Eu-ZFNMR signal.
This is difficult for our present pulsed NMR spectrometer. 

\begin{figure}[tb]
  \begin{center}
    \includegraphics[keepaspectratio=true,width=75mm]{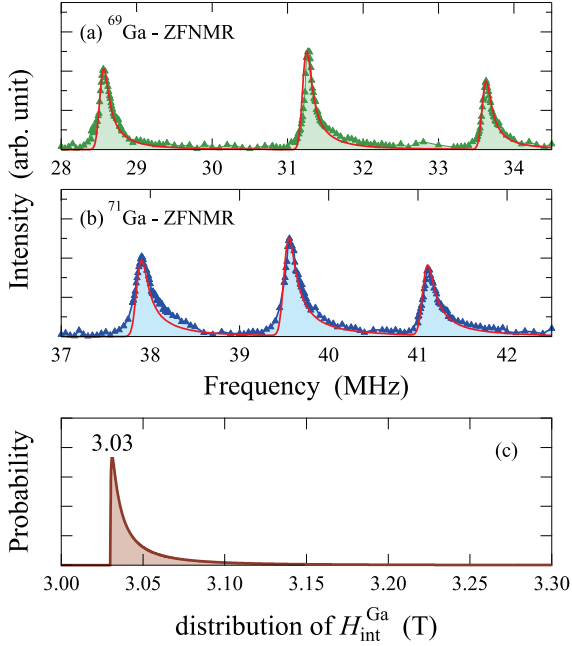}
  \end{center}
  \caption{(Color online) (a) and (b) $^{69,71}$Ga-ZFNMR spectrum at $T=4.2$ K. Solid lines indicate a calculated spectrum for $^{69,71}$Ga in the case of $H_{\rm{int}}^{\rm{Eu}}\perp V_{zz}^{\rm{Ga}}$ assuming a log-normal distribution of the internal field $H_{\rm{int}}^{\rm{Ga}}$, as shown in (c).}\label{Ga}
\end{figure}

Next, we focus on the Ga-NMR.
Ga has two isotopes, $^{69}$Ga and $^{71}$Ga, with a nuclear spin $I=3/2$.
Therefore, the spectrum consists of the center peak and two satellite peaks due to the nuclear quadrupole interaction for both Ga nuclei.
Moreover, there are two independent Ga sites in EuGa$_{4}$ crystallographically; hence, four spectra are expected in the AFM state of EuGa$_{4}$.
However, we found only two Ga-ZFNMR spectra at $T=4.2$ K, as shown in Fig. \ref{Ga}.
The local symmetries at both Ga sites give rise to an axially symmetric EFG with $V_{zz}$ parallel to the $c$-axis, as in the case of the Eu site.
Thus, the nuclear spin Hamiltonian at both Ga sites can be written as eq. (\ref{Hami}).
The inequivalent separations between the center peak and two satellite peaks indicate that $V_{zz}^{\rm{Ga}}$ is not parallel to $H_{\rm{int}}^{\rm{Ga}}$, which denotes the internal magnetic field at the Ga site.
The numerical calculation of the NMR spectrum, assuming $V_{zz}^{\rm{Ga}}\perp H_{\rm{int}}^{\rm{Ga}}$ and a log-normal distribution of $H_{\rm{int}}^{\rm{Ga}}$ as shown in Fig. \ref{Ga}(c), explains well the experimental results shown by solid lines in Figs. \ref{Ga}(a) and \ref{Ga}(b).
Although the origin of the distribution of $H_{\rm{int}}^{\rm{Ga}}$ is not known, we speculate that this relates to the distribution of the EFG at the Eu site. 
From this calculation, $H_{\rm{int}}^{\rm{Ga}} \approx 3.03$ T and $^{69}\nu_{Q}(^{71}\nu_{Q}) = 5.08(3.21)$ MHz are obtained at 4.2 K.
The same value of the internal field and the relation $^{69}\nu_{Q}/^{71}\nu_{Q}$ $\approx$ $^{69}Q/^{71}Q$ reveal that these spectra come from the same Ga site; namely, the spectra come from either the Ga1 or Ga2 site.

\begin{figure}[tb]
  \begin{center}
    \includegraphics[keepaspectratio=true,width=75mm]{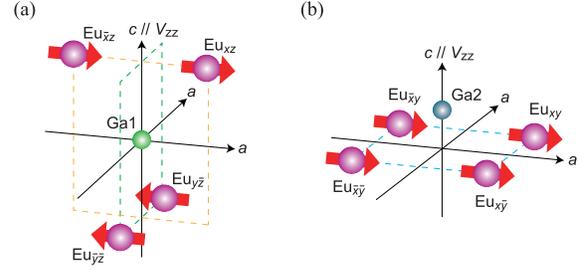}
  \end{center}
  \caption{(Color online) Coordinations of nearest-neighbor Eu sites around the (a) Ga1 and (b) Ga2 sites. The arrows on the Eu atoms indicate the magnetic moment, assuming antiferromagnetic structure type-I (see text).}\label{Ga12-site}
\end{figure}

The magnetic dipole interaction from the four nearest-neighbor Eu moments shown in Fig. \ref{Ga12-site} with $\mu_{\rm{Eu}} = 6.07$ $\mu_{\rm{B}}$\cite{Kawasaki_EuGa4} produces the magnetic dipole fields $\mib{H}_{\rm{dip}}^{\rm{Ga1}} = (0.334,0,0)$ T and $\mib{H}_{\rm{dip}}^{\rm{Ga2}} = (0.176,0,0)$ T for the Ga1 and Ga2 sites, respectively.
This is one order of magnitude smaller than $H_{\rm{int}}^{\rm{Ga}}$, indicating that the transferred hyperfine field is dominant for $H_{\rm{int}}^{\rm{Ga}}$.
Hence, we discuss the internal fields at both Ga sites, considering the short range transferred hyperfine interaction between the Ga nucleus and ordered moments on the four nearest-neighbor Eu sites.
A similar analysis was applied to discuss the magnetic structure of BaFe$_{2}$As$_{2}$.\cite{Kitagawa_BeFe2As2}

First, we focus on the internal field at the Ga1 site.
The internal field can be written as the sum of contributions from each Eu site as
\begin{equation}
\mib{H}_{\rm{int}}^{\rm{Ga1}} = \sum_{i} \mib{B}_{i}\cdot \mib{m}_{i}\quad (i = xz, \bar{x}z, y\bar{z}, \textrm{and}\ \bar{y}\bar{z}),\label{Hhyp}
\end{equation}
where $\mib{B}_{i}$ is the hyperfine coupling tensor between the Ga1 and Eu$_i$ sites, and $\mib{m}_{i}$ is the ordered moment on the Eu$_i$ site. 
From the local symmetry of the Ga1 site, the components of $\mib{B}_{i}$ can be written as
\begin{eqnarray}
\mib{B}_{xz} &=& 
\left(
\begin{array}{ccc}
B_{11} & 0 & B_{13} \\
0 & B_{22} & 0 \\
B_{31} & 0 & B_{33} 
\end{array}
\right),\nonumber\\ 
\mib{B}_{\bar{x}z} &=& 
\left(
\begin{array}{ccc}
B_{11} & 0 & -B_{13} \\
0 & B_{22} & 0 \\
-B_{31} & 0 & B_{33} 
\end{array}
\right),\nonumber\\
\mib{B}_{y\bar{z}} &=& 
\left(
\begin{array}{ccc}
B_{22} & 0 & 0 \\
0 & B_{11} & -B_{13} \\
0 & -B_{31} & B_{33} 
\end{array}
\right),\nonumber\\
\mib{B}_{\bar{y}\bar{z}} &=& 
\left(
\begin{array}{ccc}
B_{22} & 0 & 0 \\
0 & B_{11} & B_{13} \\
0 & B_{31} & B_{33} 
\end{array}
\right).
\end{eqnarray}
In the case of the type-I AFM structure shown in Fig. \ref{Ga12-site}(a), the magnetic moments are
\begin{equation}
\mib{m}_{xz} = \mib{m}_{\bar{x}z} = -\mib{m}_{y\bar{z}} = -\mib{m}_{\bar{y}\bar{z}} \equiv \mib{m}_{ab}.
\end{equation}
To include the possibility that the Eu moments are not parallel to the $a$-axis, we set $\mib{m}_{ab} = (m_a, m_b, 0)$.
Therefore,
\begin{eqnarray}
\mib{H}_{\rm{int}}^{\rm{Ga1}} &=& (\mib{B}_{xz} + \mib{B}_{\bar{x}z} - \mib{B}_{y\bar{z}} - \mib{B}_{\bar{y}\bar{z}})\cdot \mib{m}_{ab}\nonumber \\
&=&
\left(
\begin{array}{ccc}
2(B_{11} - B_{22}) & 0 & 0 \\
0 & 2(B_{22} - B_{11}) & 0 \\
0 & 0 & 0 
\end{array}
\right)
\left(
  \begin{array}{c}
    m_a \\
    m_b \\
    0 \\
  \end{array}
\right)\nonumber\\
&=& 2(B_{11} - B_{22})
\left(
  \begin{array}{c}
    m_a \\
    -m_b \\
    0 \\
  \end{array}
\right).
\end{eqnarray}
Thus, the internal field at the Ga1 site appears in the $ab$-plane.

Next, we focus on the internal field at the Ga2 site.
The magnetic moment configuration is shown in Fig. \ref{Ga12-site}(b).
The internal field at the Ga2 site is also written as eq. (\ref{Hhyp}) with a different hyperfine coupling tensor denoted $\mib{C}_{i}$ ($i = xy$, $\bar{x}y$, $x\bar{y}$, and $\bar{x}\bar{y}$).
From the local symmetry of the Ga2 site, the components of $\mib{C}_{i}$ can be written as
\begin{eqnarray}
\mib{C}_{xy} &=& 
\left(
\begin{array}{ccc}
C_{11} & C_{12} & C_{13} \\
C_{12} & C_{11} & C_{13} \\
C_{31} & C_{31} & C_{33} 
\end{array}
\right),\nonumber\\ 
\mib{C}_{\bar{x}y} &=& 
\left(
\begin{array}{ccc}
C_{11} & -C_{12} & -C_{13} \\
-C_{12} & C_{11} & C_{13} \\
-C_{31} & C_{31} & C_{33} 
\end{array}
\right),\nonumber\\
\mib{C}_{x\bar{y}} &=& 
\left(
\begin{array}{ccc}
C_{11} & -C_{12} & C_{13} \\
-C_{12} & C_{11} & -C_{13} \\
C_{31} & -C_{31} & C_{33} 
\end{array}
\right),\nonumber\\ 
\mib{C}_{\bar{x}\bar{y}} &=& 
\left(
\begin{array}{ccc}
C_{11} & C_{12} & -C_{13} \\
C_{12} & C_{11} & -C_{13} \\
-C_{31} & -C_{31} & C_{33} 
\end{array}
\right).
\end{eqnarray}
Following a similar procedure with the magnetic moments
\begin{equation}
\mib{m}_{xy} = \mib{m}_{\bar{x}y} = \mib{m}_{x\bar{y}} = \mib{m}_{\bar{x}\bar{y}} \equiv \mib{m}_{ab},
\end{equation}
we obtain
\begin{eqnarray}
\mib{H}_{\rm{int}}^{\rm{Ga2}} &=& (\mib{C}_{xy} + \mib{C}_{\bar{x}y} + \mib{C}_{x\bar{y}} + \mib{C}_{\bar{x}\bar{y}})\cdot \mib{m}_{ab}\nonumber \\
&=&
\left(
\begin{array}{ccc}
4C_{11} & 0 & 0 \\
0 & 4C_{11} & 0 \\
0 & 0 & 4C_{33} 
\end{array}
\right)
\left(
  \begin{array}{c}
    m_a \\
    m_b \\
    0 \\
  \end{array}
\right)\nonumber\\
&=& 4C_{11}
\left(
  \begin{array}{c}
    m_a \\
    m_b \\
    0 \\
  \end{array}
\right).
\end{eqnarray}
Thus, the internal field at the Ga2 site also appears in the $ab$-plane.
This result does not change even if we consider the Eu magnetic moments located above and below the Ga2 site.
The directions of the internal field at both Ga sites are in good agreement with the experimental result $H_{\rm{int}}^{\rm{Ga}}\perp c$-axis.
However, we could not clarify which site contributes to the Ga-ZFNMR signals.

\begin{table}
\caption{Comparison of the experimental values of $\nu_{Q}$ with theoretical calculation assuming paramagnetic (PM) and ferromagnetic (FM) states of EuGa$_{4}$ (see text).}
\label{calcnuQ}
\begin{center}
\begin{tabular}{ccccc}
\hline
 & $^{153}\nu_{Q}(\rm{Eu})$ & $^{69}\nu_{Q}(\rm{Ga1})$ & $^{69}\nu_{Q}(\rm{Ga2})$ \\
\hline
Calc. (PM) & 9.28 & 10.29 & 0.833 \\
Calc. (FM) & 10.9 & 11.5 & 2.8 \\
\hline
Experiment & 30.5 & \multicolumn{2}{c}{5.08} \\
\hline
\end{tabular}
\end{center}
\end{table}
For further investigation, we have calculated the nuclear quadrupole frequency based on the band calculation by a full potential linear augmented plane wave (FLAPW) method on the basis of a local density approximation (LDA) assuming paramagnetic and ferromagnetic states of EuGa$_{4}$ without spin-orbit interaction.
Here, we used the lattice parameter reported by Nakamura \textit{et al}.\cite{Nakamura_EuGa4}
The estimated values are listed in Table \ref{calcnuQ}.
These values are not in good agreement with the experimental values, especially in the case of the $^{153}$Eu nucleus.
This disagreement probably comes from the lack of a spin-orbit interaction and/or a large Coulomb interaction in $4f$ electrons beyond the LDA in the present calculation.
As for the Ga nucleus, $^{69}\nu_{Q}^{\rm LDA}(\rm{Ga2})$ for the ferromagnetic state is comparatively close to the experimentally obtained $^{69}\nu_{Q}$.
Therefore, we speculate that the observed Ga-ZFNMR spectra are derived from the Ga2 site, and the internal field at the Ga1 site is close to zero, which may be caused by almost the same value of the hyperfine coupling tensor components $B_{11}$ and $B_{22}$. 
If this speculation is true, the $^{69}$Ga nuclear quadrupole resonance signal at the Ga1 site should be observed at approximately $10\sim 20$ MHz considering the theoretical estimation of $^{69}\nu_{Q}^{\rm LDA}(\rm{Ga1})$.
However, we have not observed the signal until now, the reason for which is not clear.

\begin{figure}[tb]
  \begin{center}
    \includegraphics[keepaspectratio=true,width=80mm]{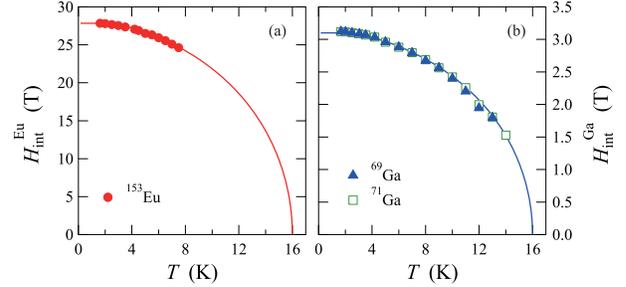}
  \end{center}
  \caption{(Color online) Temperature dependence of the internal fields at the (a) Eu site and (b) Ga site. Solid lines indicate a calculation of the Brillouin function with $J=S=7/2$.}\label{Hint}
\end{figure}

Figures \ref{Hint}(a) and \ref{Hint}(b) show the temperature dependences of the internal fields at the Eu and Ga sites, respectively.
In AFM materials, the internal field is proportional to the sublattice magnetization.
The Eu$^{2+}$ state is stable and the magnetic state is well explained by the local moment picture in EuGa$_{4}$; thus, the temperature dependence of $H_{\rm{int}}$ is well explained by the Brillouin function with $J=S=7/2$ and $T_{\rm N} = 16$ K, as shown by the solid lines in Fig. \ref{Hint}.

In summary, we have carried out zero-field NMR measurement in an antiferromagnetically ordered state of EuGa$_{4}$.
The analysis of the $^{153}$Eu and $^{69,71}$Ga ZFNMR spectra with well-resolved nuclear quadrupole splitting tells us that the magnetic moments lie in the $ab$-plane.
The huge internal field at the Eu nucleus site and the temperature dependence of the internal fields at both Eu and Ga sites reveal that a well-localized $f$-electron-derived magnetic state is realized in EuGa$_{4}$.

\section*{Acknowledgment}
The authors thank H. Shimabukuro, K. Nema, and H. Yasutomi for their experimental support and A. Nakamura for useful discussion.
This work was supported by a Grant-in-Aid for Scientific Research on Innovative Areas ``Heavy Electrons" (No. 20102007) of The Ministry of Education, Culture, Sports, Science, and Technology, Japan.

\end{document}